# A LATTICE STUDY OF THE GLUON PROPAGATOR, IN THE LANDAU GAUGE[*]

P. MARENZONI[a], G. MARTINELLI[b], N. STELLA[c], M. TESTA[b]

[a] *Dip. di Ingegneria dell'Informazione*
*Università di Parma, Viale delle Scienze, 43100 Parma, Italy*
[b] *Dip. di Fisica, Università degli Studi di Roma "La Sapienza" and*
*INFN, Sezione di Roma, P.le A. Moro 2, 00185 Rome, Italy.*
[c] *Physics Department, "The University", S09 5NH Highfield, Southampton, U.K.*

## ABSTRACT

We present the results of two high-statistics studies of the gluon propagator in the Landau gauge, at $\beta = 6.0$, on different lattice volumes. The dependence of the propagator on the momenta is well described by the expression $G(k^2) = \left[M^2 + Z \cdot k^2 (k^2/\Lambda^2)^\eta\right]^{-1}$. We obtain a precise determination of $\eta = 0.532(12)$, and verify that $M^2$ does not vanish in the infinite volume limit.

The non-perturbative investigation of the behavior of the basic fields of the QCD Lagrangian is crucial to shed light on the mechanism of confinement and can be achieved through numerical lattice computations of the Gluon Propagator[1,2] [GP].

The Euclidean GP in the Landau Gauge is:

$$D_{\mu\nu}(k) = \int d^4x \text{Tr}\langle A_\mu(x) A_\nu(0)\rangle e^{-ikx} = G(k^2)\left(\delta_{\mu\nu} - \frac{k_\mu k_\nu}{k^2}\right), \quad (1)$$

where $\mu, \nu = 1, \ldots, 4$ and the trace is intended over color indices.

Recently, there has been much effort in trying to obtain a non perturbative form for $G(k^2)$, both from analytic[3-7] and numerical[8-11] analyses. With the present studies, we investigate the non-perturbative form of $G(k^2)$ and its behaviour in the infinite volume limit.

In tab.1, the parameters of our simulations are summarized. The gauge fields have been generated with a Hybrid Monte Carlo algorithm[12]. The Landau gauge-fixing has been performed and checked carefully, being this a crucial point when dealing with gauge-dependent quantities. The fluctuation left-over after gauge-fixing $\langle \partial_\mu A_\mu(x)\rangle_{\text{Latt}} \leq 10^{-6}$, are absolutely negligible[1], with respect to the statistical errors. This can be checked since the condition $\partial_\mu A_\mu(x) = 0$ implies $\partial_t A_0(\vec{0}, t) = 0$. One can define the correlation $\langle A_0(t) A_0(0)\rangle$ at zero momentum, and study its time derivative. We have shown[1,2] that it is zero within errors, and the gluon field $A_0(\vec{0}, t)$ is constant at the level of 0.008% on each individual configuration. These results demonstrate that the present gauge-fixing procedure is the most effective, among those implemented in the literature[9-11].

Accordingly to eqn.1, we compute 2-pt functions of the gluon field, defined in term of the link variable $U_\mu(x)$ as $A_\mu(x) = [U_\mu(x) - U_\mu^\dagger(x)]/2i$, which, using spectral

---
[*]Talk presented by N.Stella.

| $\beta$ | # confs. | Volume | $\partial_\mu A_\mu$ |
|---|---|---|---|
| 6.0 | 1000 | $16^3 \times 32$ | $< 10^{-6}$ |
| 6.0 | 500 | $24^3 \times 48$ | $< 10^{-6}$ |

Table 1: Summary of the parameters of our simulations.

decomposition and translation invariance, can be written as

$$D(t,\vec{k}) = \sum_{\vec{x}} \text{Tr}\langle A_j(x)A_j(0)\rangle e^{i\vec{k}\cdot\vec{x}} = \sum_{|i\rangle} \frac{|\langle A_j(0)|i\rangle|^2}{\mathcal{N}_i} e^{-E_i t}, \qquad j = 1,\ldots,3, \qquad (2)$$

where the sum is over the states which couple to the gluon field and $E_i$ is the energy of the state $|i\rangle$. From eq. (2), the effective energy, defined as

$$\omega_{\text{eff}}(t,\vec{k}) = \log \frac{D(t,\vec{k})}{D(t+1,\vec{k})}. \qquad (3)$$

should be a decreasing function of the time, for any value of the momentum $\vec{k}$. In fig. 1, we show that $\omega_{\text{eff}}(t,\vec{k})$ is increasing with time, for all the momentum combinations considered. Hence, it is impossible[1,8] to fit the GP to a sum of single particle pole function, neither if physical states ($\mathcal{N}_i > 0$) nor "ghost" ($\mathcal{N}_i < 0$) are considered. This is an unacceptable feature for the propagator of a physical particle. To avoid systematic uncertainties due to the presence of infrared and ultraviolet cut-offs on the lattice, one can study the GP in the intermediate region of momenta. We find that $G(k^2)$ is well described by the following modeling function

$$G(k^2) = \frac{1}{M^2 + Zk^2 \left(\frac{k^2}{\Lambda^2}\right)^\eta}, \xrightarrow{\text{on the lattice}} \frac{1}{M_{\text{L}}^2 + Z_{\text{L}}(k^2)^{1+\eta}}, \qquad (4)$$

which depends on the parameters $\eta$, $M_L^2$, and $Z_L$. The stability and the quality of the fits have been checked in different ways[1,2]. On the two volumes, we find (see fig. 2)

$$V = 16^3 \times 32 \begin{cases} M_L^2 = 2.8(1) \times 10^{-3} \\ Z_L = 9.01(4) \times 10^{-2} \\ \eta = 0.56(6) \\ \chi^2_{\text{ndof}} = 1.5 \end{cases} \qquad V = 24^3 \times 48 \begin{cases} V = 24^3 \times 48 \\ M_L^2 = 4.46(9) \times 10^{-3} \\ Z_L = 0.102(1) \\ \eta = 0.532(12) \\ \chi^2_{\text{ndof}} = 1.08 \end{cases} \qquad (5)$$

We observe that the finite volume has a very small effect in the value of the anomalous dimension, provided that the range of $k^2$ is large enough. Indeed, we obtain a fairly accurate determination of $\eta$. The two determinations of $M^2$ are inconsistent, and $M^2$ increases with the volume. This feature rules out the hypothesis that the non-zero value of $M^2$ is merely due to finite volume effects. If this were the case, we would expect $M^2$ to scale roughly as $1/L^2$. It is possible to try a first, very crude extrapolation of the value of $M^2$ to the infinite volume limit. Using

$$M^2(V) = M^2(V = \infty) + \text{cost}\frac{1}{\sqrt{V}} \qquad \text{we get} \quad M^2(V = \infty) = 6.202(8) \times 10^{-3}. \qquad (6)$$

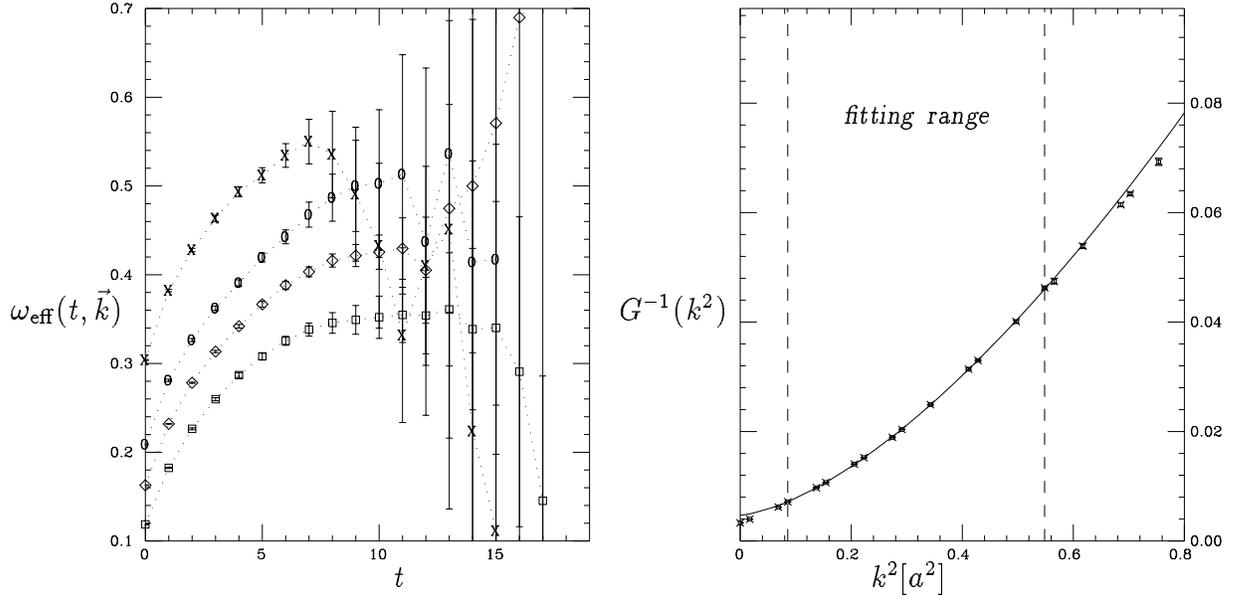

Figure 1: Effective energy for gluon 2-pt functions. The four curves correspond to the following momenta: $\square : \vec{k} = \vec{0}$, $\diamond : \vec{k} = (2\pi/24, 0, 0)$, x: $\vec{k} = (2\pi/24, 2\pi/24, 0)$ and $0 : \vec{k} = (4\pi/24, 0, 0)$. Figure 2: Best fit of the propagator in momentum space to the function 5. The figure corresponds to the case $V = 24^3 \times 48$.

Using $a^{-1} \sim 2\mathrm{GeV}$, as determined by several simulations at $\beta = 6.0$, we obtain

$$M^2_{\mathrm{phys}} \simeq (160 \text{ MeV})^2 \simeq (\Lambda_{QCD})^2. \tag{7}$$

An interpretation of this result, in connection with colour confinement is, at present, absent. However, we stress that it is fundamental to understand the behaviour of both $\eta$ and $M^2$ in the continuum limit.

**Acknowledgments**

We are indebted to the Thinking Machines Corporation for allowing us to perform these simulations. G.M. and M.T. acknowledge the partial support os MURST, Italy. N.S. thanks the Noopolis-Sovena Foundation for financial support.We are indebted to the Thinking Machines Corporation for allowing us to perform these simulations. G.M. and M.T. acknowledge the partial support os MURST, Italy. N.S. thanks the Noopolis-Sovena Foundation for financial support.